\documentclass[preprint,showpacs,preprintnumbers,amsmath,amssymb]{revtex4}

\usepackage{graphicx}% Include figure files
\usepackage{dcolumn}% Align table columns on decimal point
\usepackage{bm}% bold math

\begin{document}

%\preprint{Preprint}

\title{New Anomalies in Cosmic Microwave Background Anisotropy: Violation of the Isotropic Gaussian Hypothesis in Low-$l$ Modes}% Force line breaks with \\

\author{Shi Chun, Su}
 \email{scsu@phy.cuhk.edu.hk}
\author{M.-C., Chu}%
 \email{mcchu@phy.cuhk.edu.hk}
\affiliation{%
Department of Physics and Institute of Theoretical Physics, The Chinese University of Hong Kong, Shatin, Hong Kong
}%

\date{\today}% It is always \today, today,

\begin{abstract}
In the standard framework of cosmology, primordial density
fluctuations are assumed to have an isotropic Gaussian distribution. We search
for deviations from this assumption in the WMAP data for the low $l$ modes of
Cosmic Microwave Background Anisotropies (CMBA), by studying the directions of
the z-axis that maximize the $l=m$ modes and the resulting amplitudes of these
modes.  We find a general alignment of the directions for $l=2$ to 10 modes to
within 1/4 of the northern hemisphere. This alignment can be regarded as a
generalization of the recently discovered alignment of the $l=2$ and 3 modes
- the so-called `Axis of Evil'.  Furthermore, we find abnormally high (low)
powers in the $l=m=6$, 12 - 17 ($l=m=5$) modes; the probabilities for
having the anomalous amplitudes of the $l=m=5$, 6, 17 modes are about 0.1\%,
1\% and 1\% respectively according to the Gaussian conjecture. The
alignment and anomalous amplitudes for these low $l$ modes are very robust
against foreground contamination or different cleaning strategies, suggesting a
cosmological origin and possibly supporting a non-standard inflation.

\end{abstract}

\maketitle

Modern theories of cosmology are based on three postulates: homogeneity and
isotropy of the Universe, and primordial fluctuations of density. These three
postulates have been verified to a large extent by the observation of the
Cosmic Microwave
Background Anisotropies (CMBA) and the large scale structures of the Universe.
One usually assumes that the fluctuations are statistically isotropic and Gaussian.
Therefore, when one expands the temperature anisotropy
$\Delta T(\theta,\phi)=T(\theta,\phi)-T_0$ ($T_0\approx$ 2.73K is the mean
blackbody temperature today) of the CMBA observed today into spherical
harmonics:
\begin{eqnarray}
 \Delta T(\theta,\phi)=\sum_{l=1}^{\infty}\sum_{m=-l}^{l}a_{lm}
 Y_{lm}(\theta,\phi),
\end{eqnarray}
the $a_{lm}$'s are expected to follow a gaussian distribution among different
samples of the Last Scattering Surface (LSS) as seen from different locations
of the Universe. Unfortunately, we can observe a sample of the LSS at a particular time in a particular position only. There is inevitably a
statistical uncertainty (cosmic variance), no matter how accurate our
observation is.\\
Recently, some anomalies in the WMAP data for low $l$'s, which are found to be
inconsistent with the isotropic gaussian hypothesis (IGH) of the primordial
fluctuations, are widely studied (e.g., \cite{1,2,3,4,5,6,7}). The most famous
anomalies are the low
quadrupole amplitude $C_2$, where $(2l+1)C_l=\sum_{m=-l}^{l}|a_{lm}|^2$,
and the alignment of multipoles for $l=2$ and 3, which is known as the
`Axis of Evil' (AOE). Although some may argue that we may just be observing
a `strange' universe by accident due to the cosmic variance, many other
suggestions have been offered to explain the anomalies (e.g.,
\cite{8,9,10,11,12,13,14}). Some claim that the foreground contaminations (e.g.,
\cite{11,12}), especially those from the galactic plane or the
Virgo supercluster (towards where the alignment of $l=2$ and 3 multipoles
directs), may give rise to the anomalies. The foreground-cleaning strategies
commonly used cannot provide CMBA maps that are free of foreground
contamination residues. More interestingly, some propose that the anomalies
are true anisotropic signals from the early Universe and suggest possible
mechanisms of producing them (e.g., \cite{13,14}). It is of obvious importance and interest to distinguish among these three possibilities.\\
In this paper, we discuss several new anomalies which will make it more
difficult to attribute the
anomalies to coincidence. We also analyze how the galactic plane and other
foreground contaminations affect the anomalies and show that they most likely
do not give rise to the anomalies.\\
We analyze
\begin{eqnarray}
 r_{lm} \equiv \max_{\hat{n}}
 \left\{
 \begin{array}{c}
  \frac{|{a_{l0}}|^2}{(2l+1)C_l}, \;\; m = 0 \\
  \frac{2|{a_{lm}}|^2}{(2l+1)C_l}, \;\; m \neq 0.
 \end{array}\right.\;
 \label{eq:one}
\end{eqnarray}
Here, $r_{lm}$ measures the maximum contribution of the multipole $(l,m)$ to
the temperature power spectrum among all possible directions $\hat{n}$ of the
z-axis ($\theta=0$). We denote the direction $\hat{n}$ where $r_{lm}$ is found
as $\hat{n}_{max}$. Equation (2) provides not only a measurement of the
multipoles' alignment, but also an indication of how dominant a multipole is
over other $m$ modes of the same $l$. In \cite{6}, $r_l = \max_{m}r_{lm}$ is
studied instead of $r_{lm}$. However, the distribution of values of $r_{lm}$'s
can be different for different $m$'s of the same $l$ in general, leading to
different expectation values of $r_{lm}$. Therefore, $r_l = \max_{m}r_{lm}$ is
biased to some particular multipoles $m$ ($m\approx l/2$) which have larger
expectation values of $r_{lm}$. We therefore study $r_{lm}$ separately.\\
We analyze the 3-year full-sky WMAP Internal Linear Combination (WILC3YR) map
released by the WMAP team\cite{15}. We also generate 100,000 Monte Carlo
simulations with the IGH so as to study the statistical properties of
$r_{lm}$'s and $\hat{n}_{max}$'s. These statistics are then used to determine
if any of those $r_{lm}$'s and $\hat{n}_{max}$'s from the observed CMBA data
are anomalous under the IGH.\\
In the lower right graph of the Fig.~1, we locate the directions $\hat{n}_{max}$
for $l=m=2$ to 10, indicated by the corresponding numbers on a Mollweide
projection of the sky for the WILC3YR map.
The Galactic coordinate system is used and rotated by $180^\circ$ around the
z-axis for better view. The directions $\hat{n}_{max}$ are two-folded, i.e. a
direction $\hat{n}_{max}$ of a multipole $(l,m)$ in the northern hemisphere
must have a corresponding direction $-\hat{n}_{max}$ in the southern
hemisphere giving the same value of $r_{lm}$. Here, we show only the
$\hat{n}_{max}$'s of the multipoles in the northern hemisphere of the map.
We find from the sky map that the directions $\hat{n}_{max}$ for
$l=m=2$ to 10 concentrate within $\sim$1/5 of the solid angle of the northern
hemisphere. Furthermore, similar to the well-known alignment of the $l=m=2$ and
$l=m=3$ multipoles, there is an equally strong alignment of the $l=m=5$ and
$l=m=6$ multipoles. The probability to have such a new AOE under the IGH is
only $\sim$1\%.

\begin{figure}[htp]
\centering
\includegraphics[width=8cm, height=5cm]{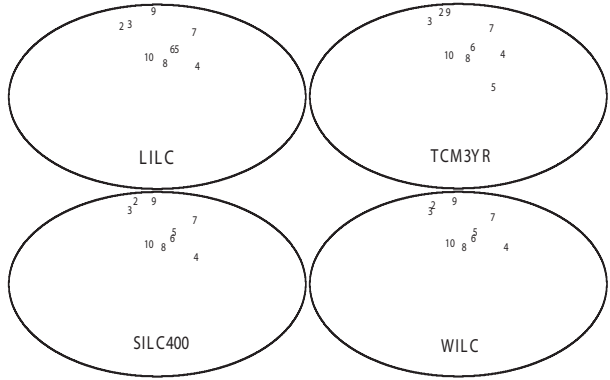}
\caption{The Mollweide projection of the full sky showing the directions
$\hat{n}_{max}$ of $l=m=2$ to 10 multipoles (indicated by the corresponding
numbers) in the four different foreground-cleaned CMBA maps: LILC, SILC400,
TCM3YR and WILC3YR map. The sky maps are rotated by $180^\circ$ around
the z-axis for better view. The directions $\hat{n}_{max}$ are two-folded with
respect to the origin. Only those $\hat{n}_{max}$'s in the northern hemisphere
are shown.}
\label{fig:}
\end{figure}

In Fig.~2, the values of $r_{ll}$'s from the WILC3YR map are shown against
$l$ by the `star' symbols while the lines indicate the median and 99\% confidence levels (solid and dash lines) of the 100,000 Monte Carlo data under
the IGH. From Fig.~2, we see that the values of $r_{ll}$'s from the WILC3YR
map show a general decreasing trend as $l$ increases. This trend is consistent
with the averaged values of the simulated $r_{ll}$'s under the IGH. Also, most
$r_{ll}$'s obtained from the WILC3YR map are consistent with these expected
values within 90\% confidence level with a few exceptions. First, we find that
$r_{55}$ obtained from the WILC3YR map is much smaller than the IGH's median
of $r_{55}$. The probability to have such a small $r_{55}$ with the IGH
is about 0.1\%. Second, another interesting but less significant anomaly is
the unusually large value of $r_{66}$. This anomaly can be reproduced by the
IGH with about 1\% chance. Third, from $l=12$ to 16, the values of $r_{ll}$
from observational data are all slightly larger than the IGH's median rather
than fluctuating around the latter. Finally, $r_{ll}$ for $l=17$ is abnormally
large, with a value that has less than 1\% probability under the IGH, and it
also aligns with the directions
$\hat{n}_{max}$ for $l=m=5$ and 6 to within $5^\circ$.\\

We study how robust these anomalies are in several ways:\\

\textbf{Different cleaning methods}\\
 We apply the same analysis to three other foreground-cleaned WMAP maps:
 LILC\cite{16}, SILC400\cite{17} and TCM3YR\cite{18}. Their directions
 $\hat{n}_{max}$ of $l=m$ multipoles are shown in Fig.~1 together with those
 of the WILC3YR map. We can see that almost all the $\hat{n}_{max}$ are
 consistent for the four CMBA maps. The only exception is the $l=m=5$ multipole
 from TCM3YR which shows a weaker alignment with the $l=m=6$ multipole.
 Nevertheless, the general alignment for $l=m=2$ to 10 multipoles is still
 strong. Even for the worst case (TCM3YR), these directions $\hat{n}_{max}$
 concentrate within $\sim$1/4 solid angle of the north hemisphere.
 The fact that a strong alignment of $l=m=5$ and 6 multipoles persists
 in the other three maps suggests that it is a true anomaly. We also
 perform the same analysis on the WILC map with the 5-year data \cite{19} (not
 shown). The directions $\hat{n}_{max}$ for all $l$'s from 2 to 10 are
 consistent with those of the 3-year data except that for $l=8$ points to a
 very different direction.\\

\begin{figure}[htp]
\centering
\includegraphics[width=8cm, height=5cm]{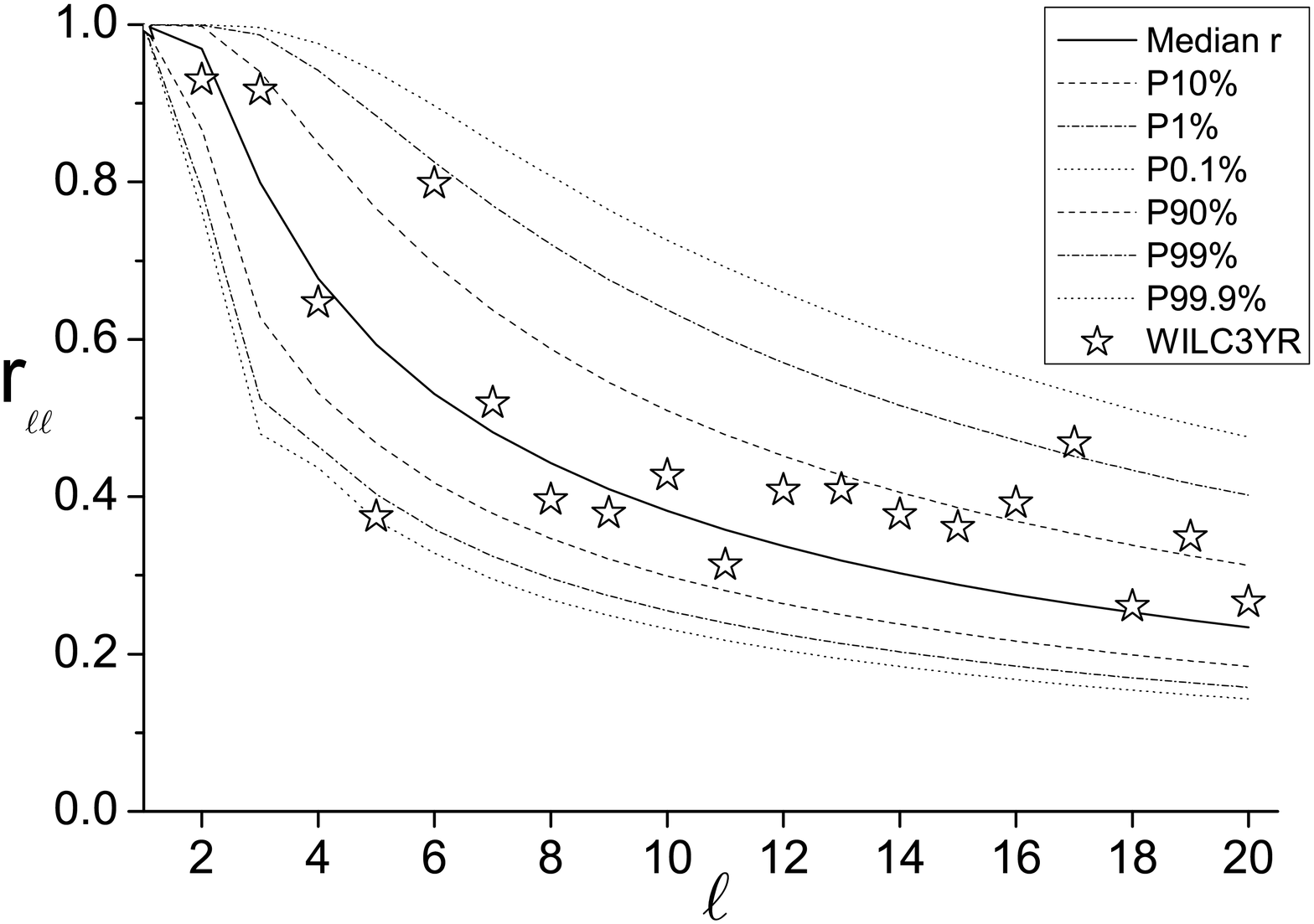}
\caption{The values of $r_{ll}$ calculated from the WILC3YR map against
$l$ are shown as stars, compared to statistical values under the IGH (lines).
$r_{55}$ ($r_{66}$) is significantly smaller (larger) than the corresponding
medians under the IGH, shown by the solid line. $r_{ll}$ for
$l=17$ is abnormally large too.}
\label{fig:}
\end{figure}

 In Fig.~3, we show the values of $r_{ll}$'s from all four foreground-cleaned
 CMBA maps. We can see that the dependence of the $r_{ll}$'s on different
 cleaning schemes are small for most $l$'s from 2 to 20. This consistency
 suggests that all anomalies mentioned above are robust features. In
 particular, the negligible dispersion of $r_{55}$ keeps the probability of
 having such a small value of $r_{55}$ as ~0.1\%. Although the value of
 $r_{66}$ from the TCM3YR map is significantly less than those of the other
 three maps, it is still well above the 90\% probability region under the IGH.
 Similarly, the anomaly of $r_{ll}$ for $l=17$ is still present for all the
 four foreground-cleaned maps. Also, the surge of $r_{ll}$ from $l=12$ to 16
 over their IGH's median persists. We remark that the dispersion of $r_{44}$ is
 exceptionally large but it is not related to any anomalies.

\begin{figure}[htp]
\centering
\includegraphics[width=8cm, height=5cm]{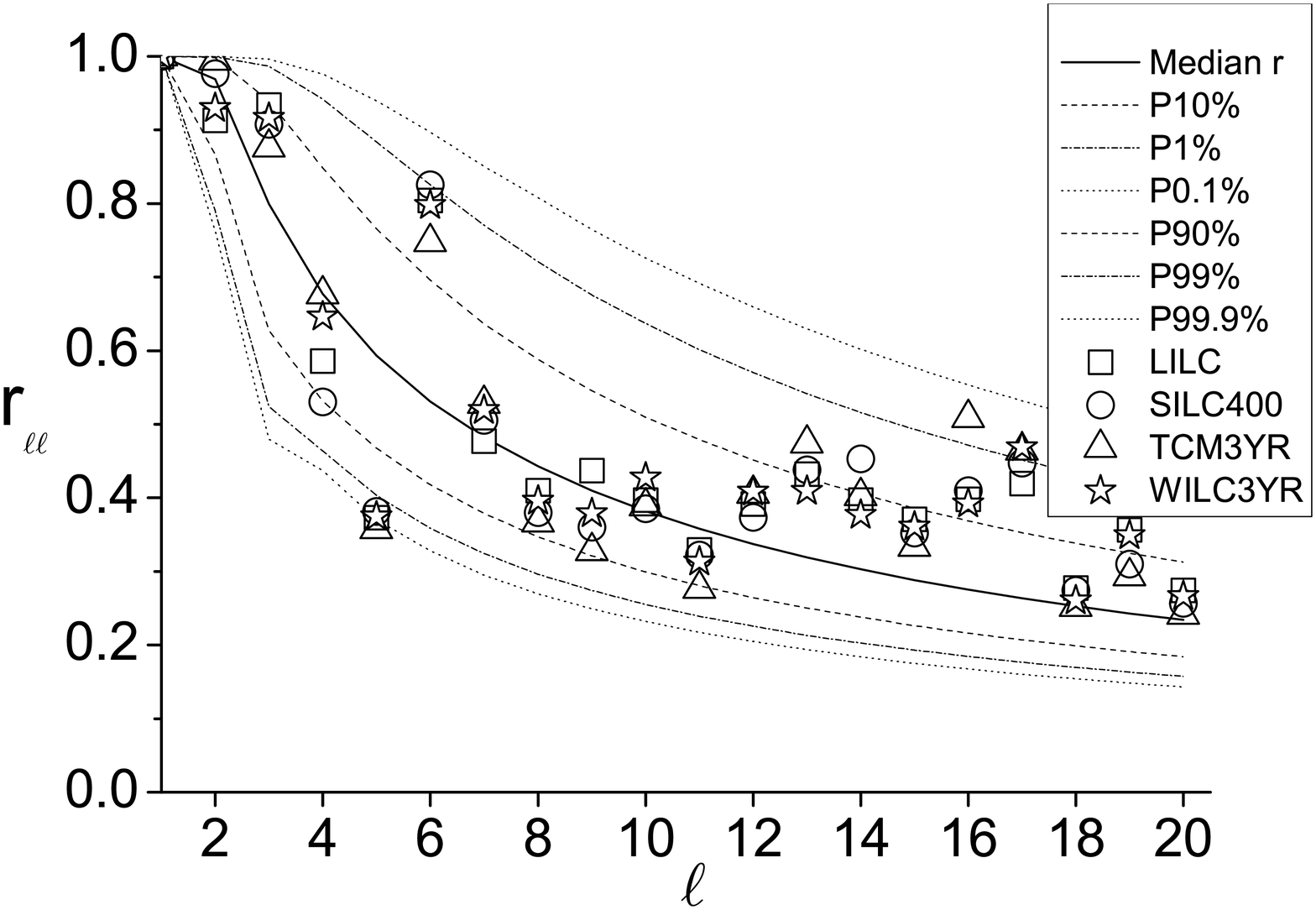}
\caption{Same as Fig.~2, but with data from all the four
foreground-cleaned CMB maps.}
\label{fig:}
\end{figure}

\textbf{Effects of the foreground contaminations}\\
 In order to examine the effects of the foreground contaminations on the
 anomalies, we introduce a simple but aggressive method: a region of the
 CMBA map is replaced by a weighted
 average of the CMBA $\Delta T$ and a simulated $\Delta T_{IGH}$ under the IGH,
 i.e. $\Delta T'=\alpha\Delta T+(1-\alpha)\Delta T_{IGH}$, where $\Delta T$
 is the observed temperature anisotropy and $\alpha$ ranges from 0 to 1.
 It is possible that the foreground
 contaminations as a deterministic effect changes the statistical properties
 of the CMBA signals under the IGH, e.g. shifting the mean values of $r_{lm}$
 and concentrating the directions $\hat{n}_{max}$. Conversely, the persistence
 of the anomalies after the replacement assures us that these foreground
 contaminations cannot independently produce the anomalies. The method is
 aggressive because it erases partially the real signals from LSS (which may
 violate the IGH itself) as well as the contaminations. If the anomalies are
 eliminated after the substitution, it is difficult, if not impossible, to
 distinguish between
 the contributions due to loss of real signals and contaminations. Therefore,
 the analysis is good at excluding contaminations as a candidate of explaining
 the anomalies but not at supporting it. The advantage of using this method
 is that it is valid even for dealing with unknown contamination residues.
 We apply the method to study the effects of the foreground contaminations on
 the galactic plane (plus point sources) and at the Virgo supercluster as
 potential sources to produce the anomalies.\\

\begin{figure}[htp]
\centering
\includegraphics[width=8cm, height=5cm]{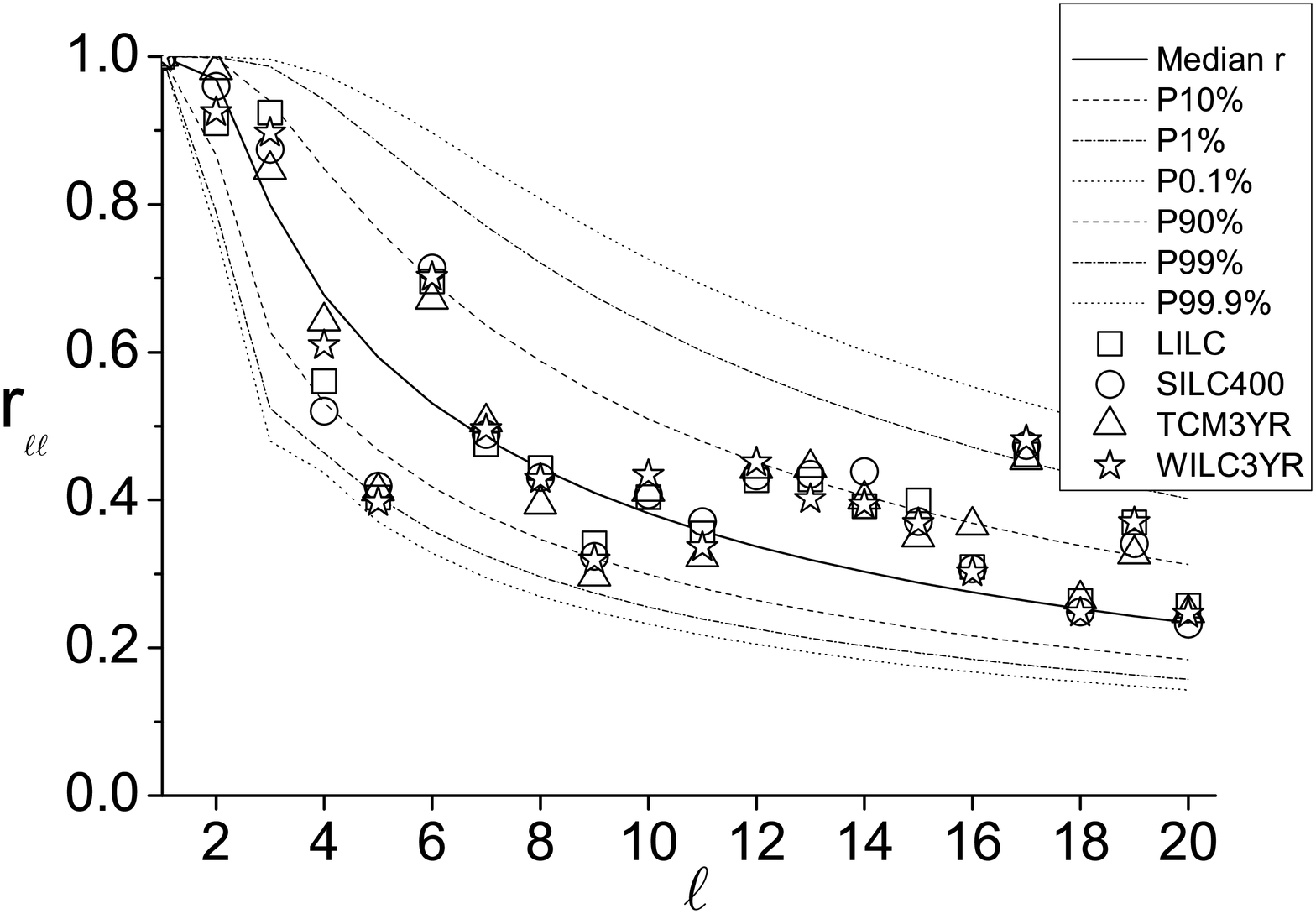}
\caption{Same as Fig.~3, but the values of $r_{lm}$'s are averages of
$r_{lm}$'s obtained from the mask-replaced maps through 1000 simulations
under the IGH with $\alpha=0.5$.}
\label{fig:}
\end{figure}

 For the galactic plane, we use the widely used standard intensity mask - kp2,
 which covers 15.3\% of the full sky including the galactic plane and point
 sources, as the region of the replacement, and $\alpha$ is set to be 0.5. The
 averaged values of $r_{ll}$'s of the
 modified maps over 1000 simulations are plotted against $l$ in Fig.~4. We can
 see that all anomalies of $r_{ll}$'s mentioned above are still significant.
 The depressed (boosted) value of $r_{55}$ ($r_{66}$) remains at about 1\%
 (10\%) probability under the IGH. The surge of values of $r_{ll}$'s from
 $l=12$ to 16 still exists
 obviously, and the value of $r_{17,17}$ is still abnormal at about 99\%
 confidence level. The directions $\hat{n}_{max}$ for $l=2,3,4,6,7$ and 10
 under replacements stay near their original directions. Although those
 for $l=5,8$ and 9 may point away to other possible directions, they
 gather in different patches respectively instead of spreading all over the
 sky. Among these patches, only one for $l=5$ and one for $l=8$ direct outsides
 the original region of the concentration or the generalized AOE.\\

 In order to test if the anomalies are caused by the foreground contamination
 at the Virgo supercluster, we replace 50\% of signals from the Virgo
 supercluster by simulated data under the IGH. The results for $r_{ll}$'s are
 shown in Fig.~5. $r_{55}$, $r_{66}$, and $r_{17,17}$ are still anomalous at
 over 99\%, 90\% and 90\% level respectively and the surge for $l=m=12$ to 16
 remains. Furthermore, such a replacement does not affect the general alignment
 of the low $l$ modes, with only the $\hat{n}_{max}$ of $l=m=8$
 pointing away from the generalized AOE for the TCM3YR map.
 We will give more details on how the $\hat{n}_{max}$'s change under foreground
 contaminations in a coming paper.\\

\begin{figure}[htp]
\centering
\includegraphics[width=8cm, height=18cm]{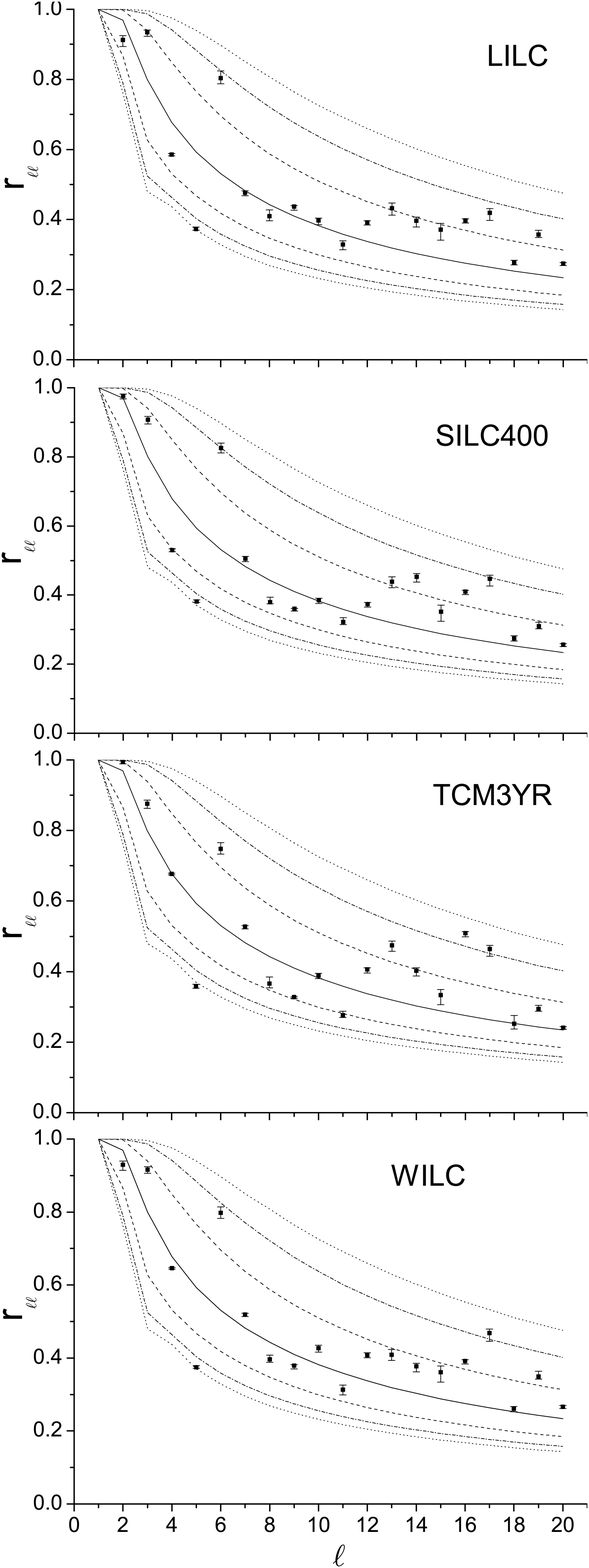}
\caption{Same as Fig.2, but for each $l$, the square indicates the original
value of $r_{ll}$. The ends of the error bar indicate the upper and lower
standard deviations of the simulated values of $r_{ll}$ (over 1,000
simulations) produced by replacing the patch of the Virgo Supercluster by the
IGH signals.}
\label{fig:}
\end{figure}

 The general alignment of the directions
 $\hat{n}_{max}$'s for $l=2$ to 10 within $\sim 1/4$ of the northern
 hemisphere is very robust and is unlikely to be eliminated by foreground
 contaminations or cleaning strategies.
 Furthermore, the anomalies of $r_{55}$, $r_{66}$, $r_{17,17}$ and the surge of
 $r_{ll}$'s for $l=12$ to 16 together are highly inconsistent with the IGH and
 robust against foreground cleanings. In fact, the values of $r_{55}$, $r_{66}$
 and $r_{17,17}$ are anomalous under the IGH at about $10^{-3}$, $10^{-2}$ and
 $10^{-2}$ levels respectively, which is extremely `strange' by coincidence.
 Our results therefore favor true cosmological origins of the anomalies and
 possibly support a non-standard inflation.\\
 Our analysis made use of HEALPIX \cite{20,21}.
 This work is supported by grants from the Research Grant Council of the
 Hong Kong Special Administrative Region, China (Project Nos. 400707 and
 400803).\\

\end{document}